\documentclass[review]{elsarticle}
\usepackage{lineno,hyperref}
\usepackage{graphicx}
\usepackage{dcolumn}
\usepackage{bm}
\usepackage{color}
\usepackage{amsmath}
\usepackage[utf8]{inputenc}
\usepackage{epstopdf}
\usepackage{booktabs}
\usepackage{epstopdf}
\usepackage{graphicx}
\usepackage{amsmath}
\usepackage{algorithm}
\usepackage{bbm}
\usepackage{amssymb}
\usepackage{algpseudocode}
\usepackage{caption}
\floatname{algorithm}{Algorithm}

\date{}
\begin{document}
\begin{frontmatter}
\title{Quantification of network structural dissimilarities based on graph embedding}
\author{
 Zhipeng Wang{\small$^{\mbox{1}}$},
  Xiu-Xiu Zhan{\small$^{\mbox{1}}$},
 Chuang Liu{\small$^{\mbox{1}}$},
 Zi-Ke Zhang{\small$^{\mbox{2,1}}$}
}


\address{$^1$Research Center for Complexity Sciences, Hangzhou Normal University, Hangzhou 311121, PR China}
\address{$^2$College of Media and International Culture, Zhejiang University, Hangzhou 310058, PR China}



\begin{abstract}
Identifying and quantifying structural dissimilarities between complex networks is a fundamental and challenging problem in network science. Previous network comparison methods are based on the structural features, such as the length of shortest path, degree and graphlet, which may only contain part of the topological information. Therefore, we propose an efficient network comparison method based on network embedding, i.e., \textit{DeepWalk}, which considers the global structural information. In detail, we calculate the distance between nodes through the vector extracted by \textit{DeepWalk} and quantify the network dissimilarity by spectral entropy based Jensen-Shannon divergences of the distribution of the node distances. Experiments on both synthetic and empirical data show that our method outperforms the baseline methods and can distinguish networks perfectly by only using the global embedding based distance distribution. In addition, we show that our method can capture network properties, e.g., average shortest path length and link density. Moreover, the experiments of modularity further implies the functionality of our method.

\end{abstract}

\end{frontmatter}

\section{Introduction}
The flexibility of network modeling and the rapid growth of network data in recent years make it urgent to design effective network comparison methods for applications such as comparing diffusion cascade of news~\cite{zhang2016dynamics}, the classification of proteins~\cite{liu2020computational} and evaluation of generative network models~\cite{hartle2020network,ali2014alignment,de2015structural}.
Researchers have proposed methods based on graph isomorphism to compare networks~\cite{zemlyachenko1985graph,kobler2012graph,babai2016graph,grohe2020graph}. The main limitations of isomorphism based methods are as follows: first of all, isomorphism based methods can only compare networks with the same size and are not scalable to large networks with millions of nodes. Secondly,  this kind of methods can only tell whether two networks are isomorphic or not but to what extent two networks are different
is hardly measured. Thanks to the mature research of  network topology mining~\cite{costa2007characterization,martinez2019comparing, tsitsulin2018netlsd,gartner2003graph}, an amount of researchers have studied how to use network characteristics, e.g., adjacency matrix, node degree, shortest path distance, to compare networks with huge and different sizes. For instance, Saxena et al.~\cite{saxena2019identifying} introduced a network similarity method based on hierarchical diagram decomposition via using canberra distance, which considers both local and global network topology. Lu et al.~\cite{lu2014complex} proposed a manifold diffusion method based on random walk, which can not only distinguish networks with different degree distributions but also can discriminate networks with the same degree distribution. Beyond the direct comparison of network topology, we have witnessed the effectiveness of using quantum information science, i.e., information entropy, in network comparison.
 For example, De Domenico et al.~\cite{de2016spectral} proposed a set of information theory tools for network comparison based on spectral entropy. Schieber et al.~\cite{schieber2017quantification} quantified the dissimilarities between networks by considering the probability distribution of the shortest path distance between nodes. Chen et al.~\cite{chen2018complex} proposed a comparison method based on node communicability and spectral entropy. The basic idea behind this kind of method is that one specific network property, such as the shortest path distance  \cite{schieber2017quantification} and node communicability matrix~\cite{chen2018complex} , is chosen to measure the information content of a network via a proper entropy. Therefore, the dissimilarity between two networks is given by the difference between the information content of them. However, we claim that the selection of one specific property as a representative of network information content may not be able to capture the information of a whole network.

Network embedding, which aims to embed each node into a low dimensional vector by preserving the network structure as much as possible, has been widely used to solve many problems in network science, e.g., link prediction~\cite{bu2019link,grover2016node2vec}, community detection~\cite{jin2019incorporating,li2016discriminative,yang2016revisiting} and network reconstruction~\cite{pio2020exploiting,xu2020variational,goyal2018graph}. In this paper, we further widen the application of network embedding, i.e., we explore how to use network embedding to characterize the dissimilarity of two networks in a state-of-the-art way. We start from using a simple and fast network embedding algorithm, i.e., \textit{DeepWalk}, to measure the distance between two nodes. The information content of a network, i.e., network similarity heterogeneity, is defined based on the node distance distribution and Jensen-Shannon divergence. Accordingly, the dissimilarity between two networks is further defined upon network similarity heterogeneity. We validate the effectiveness of the embedding based network comparison method on both synthetic and empirical networks. Compared to the baseline methods, embedding based network comparison shows high distinguishability.

The rest of the paper is organized as follows. In Section 2, we give a detailed description of the network embedding based comparison method as well as the baselines. In Section 3, we evaluate our method on various networks, including networks generated by network models, such as Watts–Strogatz model (WS) and Barab\'{a}si–Albert (BA) model, and empirical networks extracted from different domains.
In Section 4, we give the conclusion and provide possible directions for future work.

\section{Method}\label{method}

\subsection{Embedding based network dissimilarity}
Given a network $G = (V, E)$, in which $V$ represents the node set, and $E$ = $\left\{  (v_i, v_j), v_i, v_j \in V \right\}$ is the edge set. The number of nodes is given by $N$ = $\left|V\right|$, where $|*|$ indicates the cardinal number of a set. The adjacency matrix of $G$ is given by $A_{N\times N}$, in which $A_{ij}=1$ if there is a link between node $v_i$ and $v_j$, otherwise $A_{ij}=0$. We use \textit{DeepWalk} to learn the embedding vector of every node~\cite{perozzi2014deepwalk}. Concretely speaking, \textit{DeepWalk} conducts a uniform random walk to obtain node sequences as the input for a learning model, i.e., \textit{SkipGram}. The embedding vectors of the nodes contain the structure information of the original network. For a node $v_i$, we use $\vec{V_i}=\left(v_{i1}, v_{i2}, \cdots, v_{id}\right)$  to represent the embedding vector obtained from \textit{DeepWalk}. Therefore, we can define the Euclidean distance between two arbitrary nodes $v_i$ and $v_j$ as $b_{ij} $=$ \sqrt{\sum_{z=1}^d(v_{iz}-v_{jz})^2}$. Smaller $b_{ij}$ indicates that $v_i$ and $v_j$ are more similar. The Euclidean distance matrix is denoted as $B_{N \times N}$, in which $B(i,:)$ is the Euclidean distance between node $v_i$ and all the $N$ nodes. Hence we have $B(i,i)=0$. We define $B_{\max}=\max_{i,j}B_{ij}$ and $B_{\min}=\min_{i,j}B_{ij}=0$. We use $H_i=[H_{i1}, H_{i2}, \cdots, H_{iL}]$ to represent the Euclidean distance distribution of node $v_i$, in which $H_{iz}$ is the
probability that the Euclidean distance between a node and node $v_i$ follows in the bin $[B_{\min}+(z-1)\frac{B_{\max}-B_{\min}}{L}, B_{\min}+z\frac{B_{\max}-B_{\min}}{L}]$. $L$ is a tunable parameter.

We introduce Jensen-Shannon divergence to define the network dissimilarity based on the Euclidean distance distribution.
The Euclidean distance distribution heterogeneity of a network, i.e., $JS(G)$, is defined as:
\begin{equation}\label{JS}
        JS(G)=\frac{J(H_1,...,H_N)}{\log(N+1)},
\end{equation}
where $J(H_1,...,H_N)=\frac{1}{N}\sum_{i,j}H_i(j)log\frac{H_i(j)}{\mu_j}$ represents the Jensen-Shannon divergence of the node Euclidean  distance distribution and $\mu_j = \frac{\sum_{i=1}^NH_i(j)}{N}$ represents the average value of the $j_{th}$ dimension of the distribution $H$. We use $\mu=\{\mu_1, \mu_2, \cdots, \mu_L\}$ to represent the average  Euclidean distance distribution of a network.

Given two networks $G_1 = (V_1, E_1)$ and $G_2 = (V_2, E_2)$, we denote $\mu_{G_1}$ and $\mu_{G_2}$ as the average  Euclidean distance distributions of $G_1$ and $G_2$, respectively. The dissimilarity between $G_1$ and $G_2$ ($D_{NE}(G_{1},G_{2})$) is given by

\begin{equation}\label{D-NE}
       D_{NE}(G_{1},G_{2})=\omega\sqrt{\frac{J(\mu_{G_1},\mu_{G_2})}{\log2}}+(1-\omega)|\sqrt{JS({G_1})}-\sqrt{JS({G_2})}|,
\end{equation}
where ${\omega}$ $\in$ $[0,1]$ is a tunable parameter and we have $D_{NE}$ $\in$ $[0,1]$. The first term in Eq.~(\ref{D-NE}) compares the average Euclidean distance distribution of the two networks. And the second term evaluates the dissimilarity of Euclidean distance heterogeneity between the two networks.  Smaller value of $D_{NE}(G_{1},G_{2})$ indicates that $G_1$ and $G_2$ are more similar.

To obtain node embedding vector from \textit{DeepWalk}, we set the parameters such as embedding dimension $d=128$, number of walks per node $s=10$, the length of each walk $l=60$ and the context window size $w=8$.  In addition, we set $L=10$ in the Euclidean distance distribution $H_i$ $(i=1,2,\cdots, N)$. The influence of $L$ on the performance of network comparison is further given in Figure~S2-S5 in \textbf{Supplementary Note $3$.}

\begin{figure}[!ht]
\centering
\includegraphics[width=17cm,height=8cm]{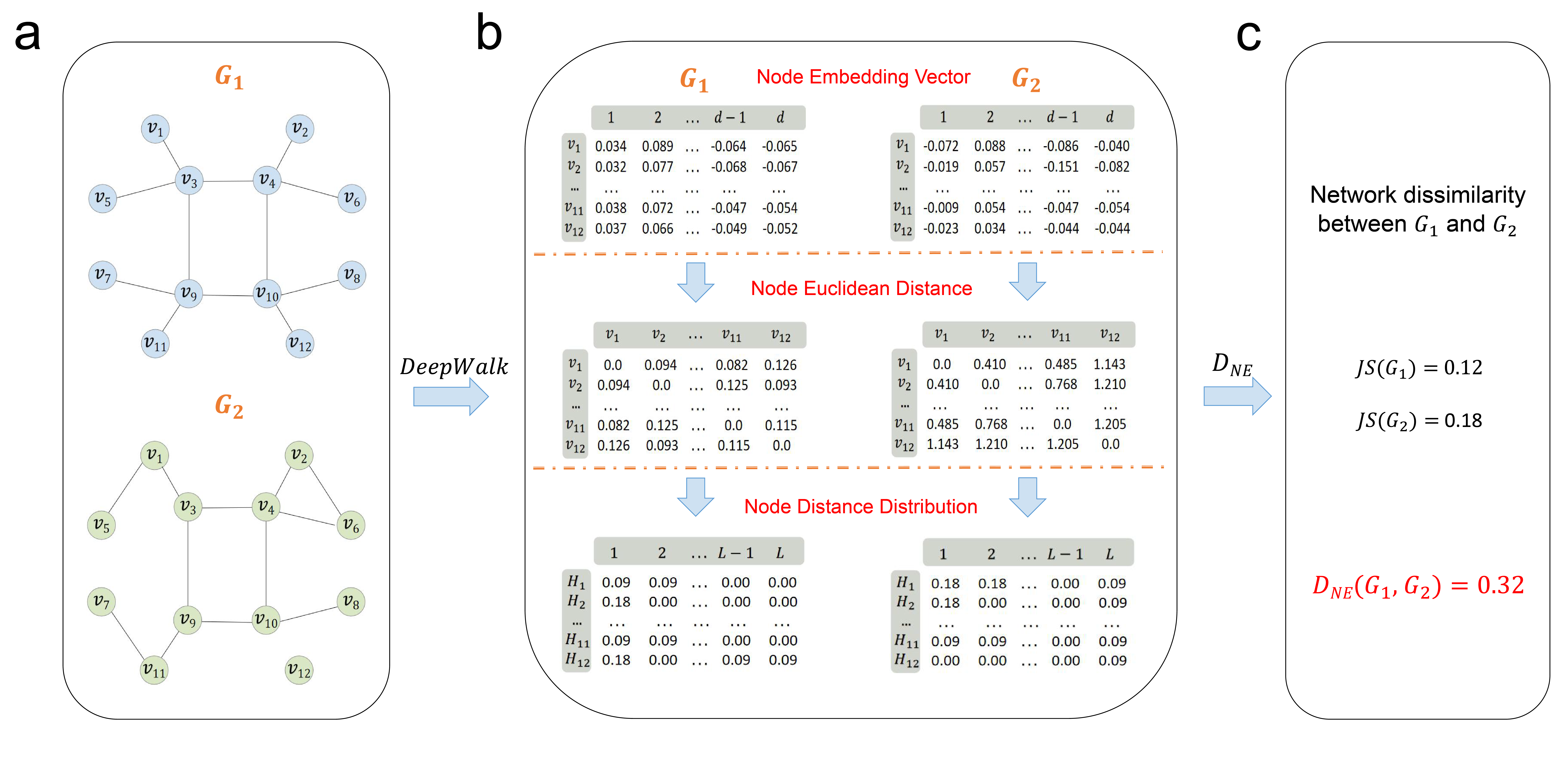}
\caption{\textbf{Illustration of embedding based network comparison method}. (a) Visualization of two network $G_1$ and $G_2$, each with 12 nodes and 12 edges. (b-c) Example of how to compute embedding based network dissimilarity, including the characterization of the node embedding, calculation of the node Euclidean distance, node distance distribution and the dissimilarity between network $G_1$ and $G_2$, where we use $\omega = 0.5$.}
\label{1}
\end{figure}

In Figure~\ref{1}, we show the network dissimilarity comparison process of our method $D_{NE}$. In Figure~\ref{1}a, we show two networks, i.e., $G_1$ and $G_2$, in which $G_1$ is a fully connected network and $G_2$ has one isolated node. The detail calculation process of the method is shown in Figure~\ref{1}b and Figure~\ref{1}c, which include the calculation of node embedding vector, node Euclidean distance matrix, node distance distribution, and the dissimilarity between the two networks via Eq. (\ref{JS}) and (\ref{D-NE}). The dissimilarity between $G_1$ and $G_2$ via $D_{NE}$ is as high as $0.32$.

\subsection{Baselines}

\textbf{Network dissimilarity based on shortest path distance distribution}.
Suppose the shortest path distance distribution of node $v_i$ is denoted by $P_i=\left\{p_i(j)\right\}$, in which $p_i(j)$ is defined as the fraction of nodes at distance $j$ from node $v_i$. Network node dispersion $NND$ measures the network connectivity  heterogeneity in terms of shortest path distance and is defined by the following equation:
\begin{equation}
        NND(G)=\frac{J(P_1,...,P_N)}{\log(dia+1)},
\end{equation}
where $dia$ represents the network diameter and $J(P_1,...,P_N)$ is the Jensen-Shannon divergence of the node distance distribution. The dissimilarity measure $D_{SP}$ is based on three distance-based probability distribution function vectors and is defined as follows:
\begin{equation}\label{D-sp}
       D_{SP}(G_{1},G_{2})=\omega_1\sqrt{\frac{J(\mu_{G_1},\mu_{G_2})}{\log2}}+\omega_2|\sqrt{NND({G_1})}-\sqrt{NND({G_2})}|+\frac{\omega_3}{2}\left( \sqrt{\frac{J(P_{\alpha G_1},P_{\alpha G_2})}{\log2}} + \sqrt{\frac{J(P_{\alpha G_1^c},P_{\alpha G_2^c})}{\log2}} \right),
\end{equation}

 where $\omega_1$, $\omega_2$, $\omega_3$, and $\alpha$ are tunable parameters, in which $\omega_1+\omega_2+\omega_3=1$. The first term in Eq.~(\ref{D-sp}) indicates the dissimilarity characterized by the averaged shortest path distance distributions, i.e.,  $\mu_{G_{1}}$ and $\mu_{G_{2}}$. The second term characterizes the difference of network node dispersion. The last term is the difference of the  $\alpha$-centrality distributions, in which $G^c$ is the complementary graph of $G$.  We set $\omega_1=\omega_2=0.45$ and  $\omega_3=0.1$, which are the default settings used in \cite{schieber2017quantification}.

\textbf{Network dissimilarity based on communicability sequence}.
The communicability matrix $C$ measures the communicability between nodes and is defined as follows:

\begin{equation}
       C=e^A=\sum_{z=0}^\infty \frac{1}{z!}A^z=\begin{Bmatrix} c_{11}&c_{12}&\cdots&c_{1N} \\
       c_{21}&c_{22}&\cdots&c_{2N} \\
       \vdots&\vdots&\ddots&\vdots\\
       c_{N1}&c_{N2}&\cdots&c_{NN}\end{Bmatrix},
\end{equation}
where $c_{ij}$ unveils the communicability between node $v_i$ and $v_j$.
 Let $P=\left\{P_1,P_2,\cdots,P_M\right\}$ be the normalized communicability sequence, in which $P_z=\frac{c_{ij}}{\sum_{i=1}^{N}\sum_{j=i}^{N}c_{ij}}$ ($1\leq z\leq M$, $1\leq i\leq j\leq N$ and $M=\frac{N(N+1)}{2}$). The Jensen-Shannon entropy of the sequence is expressed as follows:

\begin{equation}
       S(P)=-\sum_{i=1}^MP_i\log_2P_i
\end{equation}

Given two networks $G_1$ and $G_2$,  normalized communicability sequences are given by $P^{G_1}$ and $P^{G_2}$, respectively. We sort the values in  $P^{G_1}$ ($P^{G_2}$) in an ascending order and obtain new communicability sequences as $\tilde{P}^{G_1}$ ($\tilde{P}^{G_2}$). Therefore, the communicability based dissimilarity is defined as $D_{C}(G_{1},G_{2})$:
\begin{equation}
       D_{C}(G_{1},G_{2})=S\left(\frac{\tilde{P}^{G_1}+\tilde{P}^{G_2}}{2}\right)-\frac{1}{2}\left[S(\tilde{P}^{G_1})+S(\tilde{P}^{G_2})\right]
\end{equation}

\section{Results}
\subsection{Synthetic network comparison}

\begin{figure}[!ht]
\centering
\includegraphics[width=18.5cm,height=11.8cm]{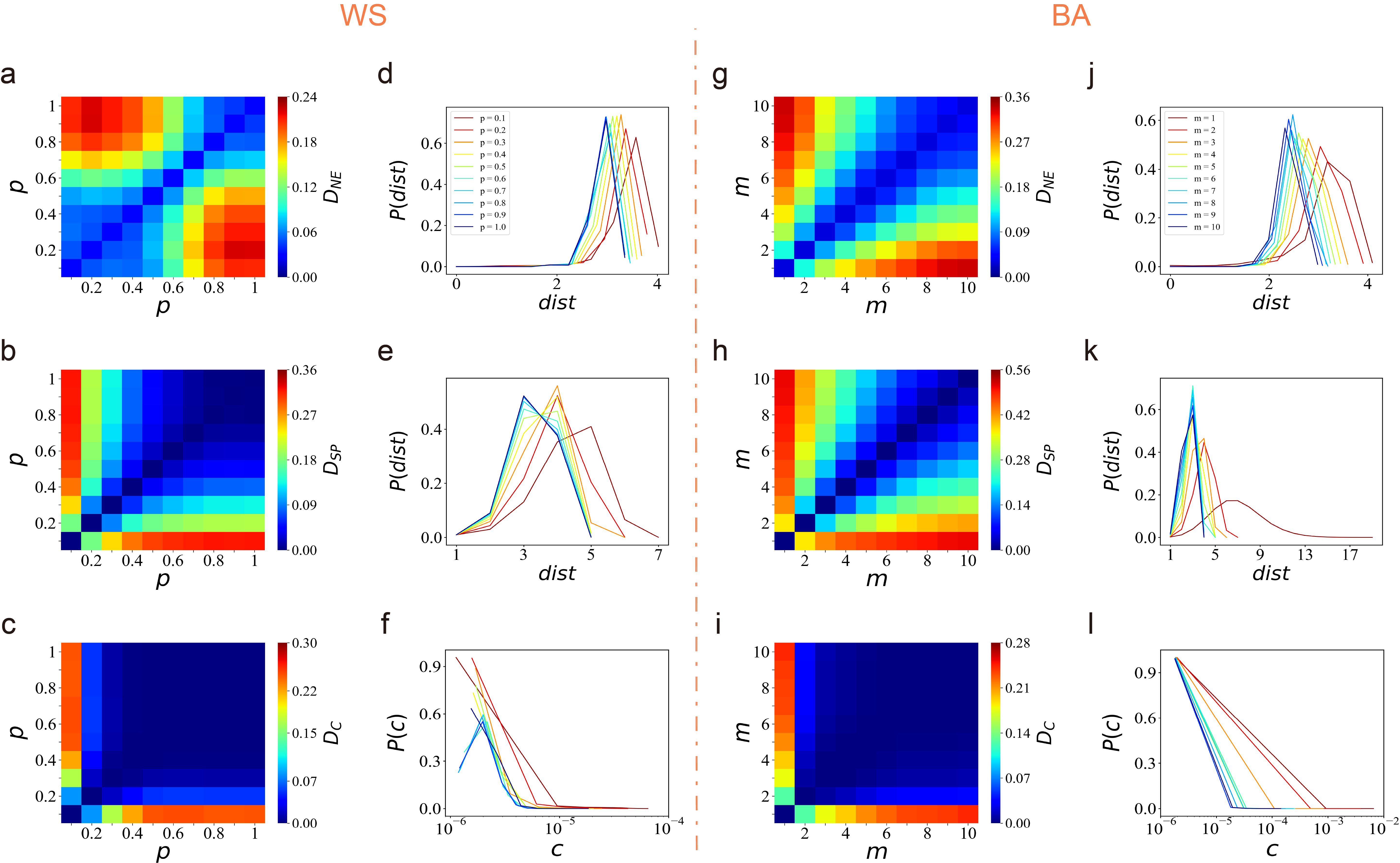}
\caption{\textbf{Performance of three comparison methods on synthetic networks}. (a-c) Dissimilarity values $D_{NE}$, $D_{SP}$, $D_{C}$ for each pair of networks generated in the WS model, respectively. (d) The embedding based average distance distributions of WS networks with different $p$ based on $D_{NE}$. (e) The average distance distributions of WS networks with different $p$ based on $D_{SP}$. (f) The node communicability distributions of WS networks with different $p$ based on $D_{C}$. (g-i) Dissimilarity values $D_{NE}$, $D_{SP}$, $D_{C}$ for each pair of BA networks under different $m$, respectively, in which $m\in\{1,2,3,4,5,6,7,8,9,10\}$. (j) The embedding based average distance distributions of BA networks with different $m$ based on method $D_{NE}$. (k) The average distance distribution on BA networks with different  $m$ based on method $D_{SP}$. (l) The node communicability distribution on BA networks with different $m$ based on method $D_{C}$. All the results are averaged over 100 realizations.}
\label{3}
\end{figure}

To verify the ability of our method in quantifying the network dissimilarity, we perform the comparison on synthetic networks including the Small-World network generated from the WS model and the Scale-Free network generated from the BA model. In all the network models, we use the network size $N=1000$. In WS model, we compare networks generated by different rewiring probability $p$, where the network average degree is 10. Figure~\ref{3}a-\ref{3}c show the dissimilarity values obtained by $D_{NE}$, $D_{SP}$, $D_{C}$ between WS networks with different $p$. Generally, we find that all the three kinds of dissimilarity values of the networks generated with similar $p$ values are much smaller than those of the networks generated with dramatically different values of $p$.
The proposed method $D_{NE}$ can detect the network dissimilarity for all the $p$ values (Figure~\ref{3}a), while $D_{SP}$ and $D_C$ can not identify the difference between networks for large values of $p$ (Figure~\ref{3}b and~\ref{3}c).  The definition of $D_{NE}$, $D_{SP}$ and $D_C$ are based on the embedding based distance distribution, the shortest path distance distribution and the node communicability distribution, respectively. The embedding based distance distributions are distinguishable across different $p$ (Figure~\ref{3}d). However, the distributions of shortest path distance and node communicability are so narrow for large $p$ values (Figure~\ref{3}e and \ref{3}f), leading to the no difference for the corresponding network comparison methods. In BA model, we generate networks by changing the value of $m$, which is the number of edges per node added at each time step. Figure~\ref{3}g-\ref{3}i show the comparison of BA networks with $m\in[1, 10]$ via the three methods. Similarly as the WS network,  $D_{NE}$ shows the best performance. The reason that $D_{SP}$ and $D_C$ perform worse is given by  the average shortest path distance distributions and the node communicability distributions when changing $m$ in Figure~\ref{3}k and \ref{3}l, respectively.

The robustness of the embedding based network comparison method is given by the parameter analysis in Figure~S1a and Figure~S1b in \textbf{Supplementary Note 1}. In all the networks, we keep the average degree as $10$. Each point in Figure~S1a shows the dissimilarity between a WS network with size $N=1000$ and $N=\{1500,2000,2500,3000,3500,4000,\linebreak4500,5000\}$, respectively. We set rewiring probability $p=0.3$. Different curves show the dissimilarity when we use different $\omega$. We find that WS network with size $N=1000$ is more similar to networks generated with close size, and different $\omega$ does not affect the similarity trend. However, large value of $\omega$ results in larger dissimilarity values between networks.
In Figure~S1b, we give the same analysis for BA model, which shows the similar results as those of WS model. We also compare the differences between the following networks: BA, WSL (it is obtained by rewiring 1$\%$ of edges in K-regular network) and WSH (it is obtained by rewiring 10$\%$ of edges in K-regular network) in Figure~S1c in \textbf{Supplementary Note 1}. Figure~S1c shows the change of the dissimilarity values with the increase of $\omega$, and the results show that large value of $\omega$ gives large dissimilarity value.  Furthermore, when $\omega=0$, indicating that only local structural information is used (Eq.~(\ref{D-NE})), the differences between the three pairs of synthetic networks are not effectively distinguished. On the contrary, when $\omega=1$, the global information of the network can better distinguish the network differences. Therefore, we set $\omega=1$ in the following analysis.

\begin{figure}[!ht]
\centering
\includegraphics[width=16cm,height=7cm]{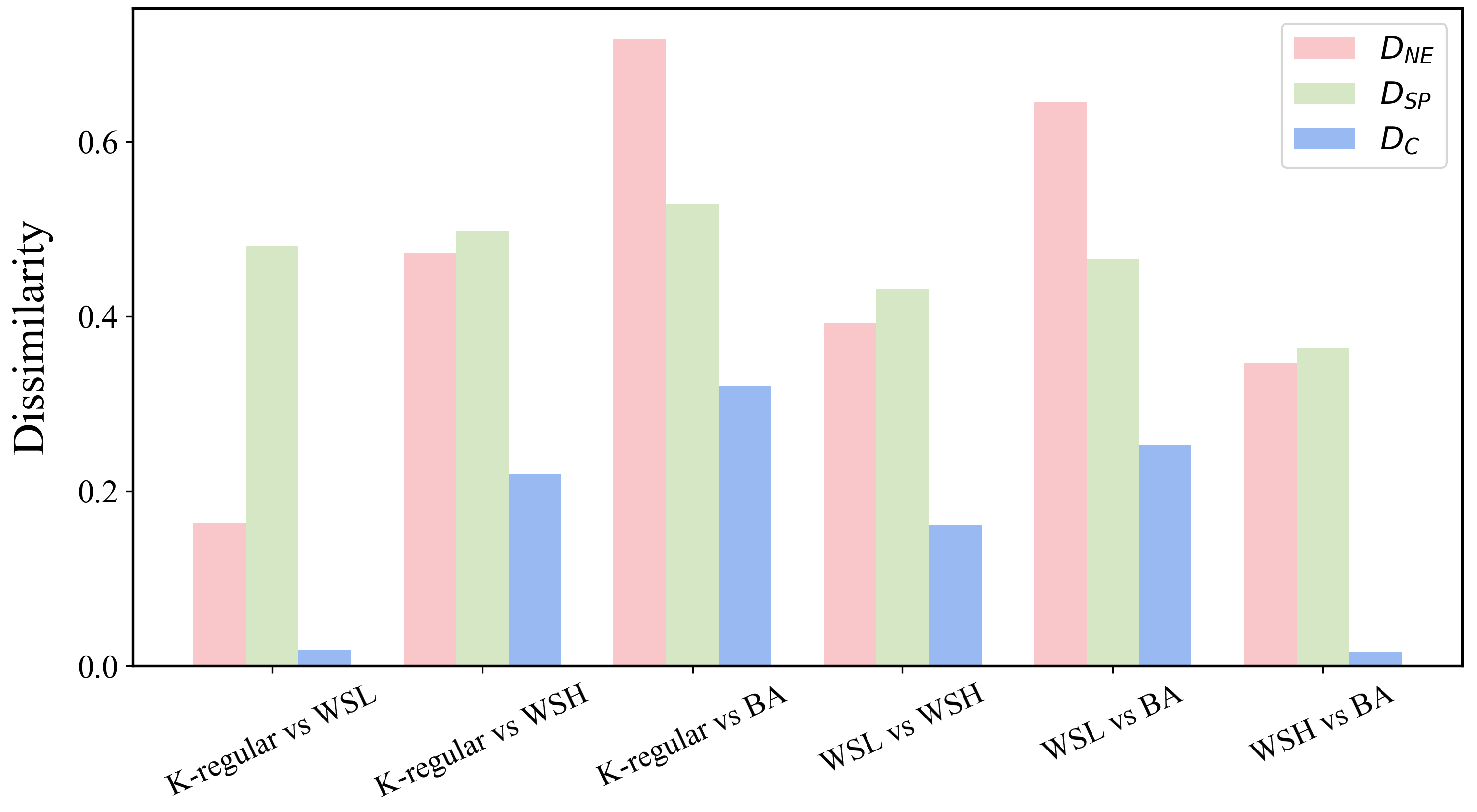}
\caption{Comparison of four synthetic networks (K-regular, WSL, WSH, and BA) using three different methods, i.e., $D_{NE}$, $D_{SP}$ and $D_{C}$, respectively , in which $D_{NE}$ is the method of network embedding, $D_{SP}$ calculates the dissimilarity value based on the method using distance distribution and $D_C$ calculates the dissimilarity value based on the communication sequence entropy. We consider network size $N = 1000$, average node degree $10$. All the results are averaged over 100 realizations with $\omega=1.0$.}
\label{2}
\end{figure}

 To compare with different dissimilarity methods, we also show the dissimilarity between four synthetic networks with the same node size $N=1000$, edge size $|E|=5000$ and average node degree $10$. The four networks include K-regular, WSL, WSH and BA network. From the generation model, we know that the descending order of similarity value between K-regular and the other three networks is as follows: WSL, WSH and BA. Figure~\ref{2} gives the dissimilarity between the four networks with three methods, i.e., $D_{NE}$, $D_{SP}$ and $D_{C}$. Figure~\ref{2} implies that dissimilarities between the four networks obtained by the three network comparison methods are consistent with the rules of network generation models. However, the dissimilarity values ($D_{SP}$) between K-regular and WSL, K-regular and WSH are almost the same and the dissimilarity values ($D_{SP}$) between the four synthetic networks are very close, indicating that the method $D_{SP}$ can not effectively discriminate the differences between these synthetic networks.

\subsection{Real networks comparison}

We validate the effectiveness of our embedding based network comparison method upon real networks from different domains. Table~\ref{tab2} gives the basic properties of the real networks, including the number of nodes ($N$), the number of edges ($|E|$), average degree ($Ad$), average path length ($Avl$), link density ($Ld$), clustering coefficient ($C$) and diameter ($dia$).  The $12$ real networks range from the protein-protein interaction (Yeast) and metabolic network (Metabolic) in biology, to the human contact (Infectious, Windsurfers), and to the social  communication networks (Pgp, Rovira, Petster, Petsterc and Irvine). The detailed descriptions of networks are given in \textbf{Supplementary Note 2}.

\begin{figure}[!ht]
\centering
\includegraphics[width=17cm,height=7cm]{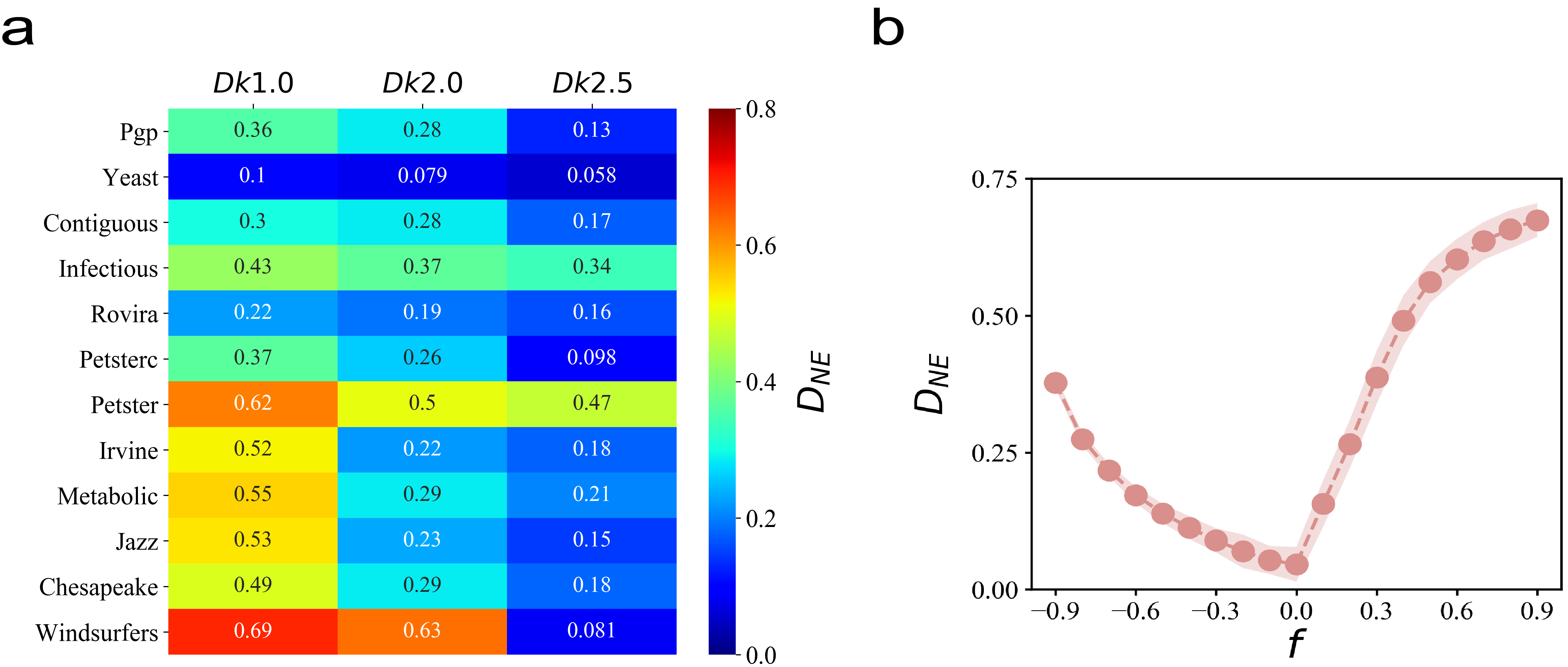}
\caption{Dissimilarity between real networks. (a) Comparison between real networks and their null models. We consider the $Dk$ models with different $k$-values ($1.0$, $2.0$, and $2.5$). (b) Dissimilarity between Petsterc network and the networks generated after certain perturbations, where negative value of $f$ corresponds to the random deletion of edges with the given ratios, and vice verse. Each point in the figure is averaged over 100 times. The shaded error area shows the standard deviation of 100 times.}
\label{4}
\end{figure}

Firstly, we show the difference between a real network and its corresponding null models in Figure~\ref{4}a. For a network $G$, we consider three kinds of null models ($k$-order null models, including $k$=1.0, 2.0 and 2.5)~\cite{orsini2015quantifying}, which is defined as $Dk1.0$, $Dk2.0$ and $Dk2.5$, respectively. Specifically, different values of $k$ indicate the preservation of network topology to different degrees. $k=1.0$ indicates that the generated network retains the degree sequence. When $k=2.0$, the degree sequence and degree correlation properties are invariant during the rewiring process. $k=2.5$ preserves the clustering spectrum property of the original network. The dissimilarity values are averaged over $100$ repeated independent runs. With the increase of $k$, the dissimilarity between a real network and its randomized networks tends to be smaller across different networks (each row in Figure~\ref{4}a). The pattern of the network dissimilarity values is consistent with the randomization process, where larger $k$ indicates that the randomized networks share more properties with the original network, leading to the more similarity to the original network.

\begin{table}[!ht]
\begin{center}
\setlength{\abovecaptionskip}{0.2cm}
\setlength{\belowcaptionskip}{0.1cm}
\caption{Basic properties of real networks, in which $N$, $|E|$, $Ad$, $Avl$, $Ld$, $C$ and $dia$ represent the number of nodes, the number of edges, average degree, average shortest path length, link density, average clustering coefficient and network diameter, respectively.}
\label{tab2}
\begin{tabular}{cccccccc}
\toprule
\textbf{Networks} & \textbf{$N$} & \textbf{$|E|$}  & $Ad$ & \textbf{$Avl$}  & $Ld$ & $C$ & \textbf{$dia$}\\ \hline
\midrule
Pgp &   10,680 & 24,316 &  4.55 & 7.49 & 0.0004 & 0.266& 24 \\
Yeast &   1,870 &  2,203  & 2.44 &  6.81 & 0.0013 & 0.067 & 19 \\
Contiguous &   49 & 107 &  4.37 & 4.16 &  0.0910 & 0.497 & 11 \\
Infectious &    410 & 2,765 & 13.49 &3.63 &  0.0330 & 0.456 & 9 \\
Rovira &   1,133 &  5,451 &  9.62 & 3.61 & 0.0085 & 0.220 & 8 \\
Petsterc &   2,426 & 16,631&  13.71 & 3.59 & 0.0057 & 0.538 & 10 \\
Petster &   1,858 & 12,534 & 13.49 & 3.45 &0.0073 & 0.141 & 14 \\
Irvine &   1,899 & 59,835 &  14.57 & 3.06& 0.0079  & 0.109 & 8 \\
Metabolic &   453 & 2,025 & 8.94 & 2.68 & 0.0198 & 0.646 & 7 \\
Jazz &   198 &  2742 &  27.69 & 2.24 & 0.1406 & 0.617 & 6 \\
Chesapeake &   39 & 170 & 8.72 & 1.83 & 0.2294 & 0.450 & 3 \\
Windsurfers &   43 & 336 & 15.63 & 1.69 & 0.3721 & 0.653 & 3 \\

\bottomrule
\end{tabular}
\end{center}
\end{table}

We compare the real networks with the networks after certain perturbation. The perturbation is performed as follows: for a given network, we randomly add (or delete) a certain fraction $f\in[0, 1]$ of edges, and then compare the dissimilarity between the original network and the perturbed network.
Positive $f$ represents addition process, and negative $f$ represents deletion process. Figure~\ref{4}b shows the dissimilarity between Petsterc network and the perturbed networks after random addition or deletion of edges. It implies that the more we perturb the network, the more dissimilar it is to the original network. We show the similar trend of the other networks in Figure~S3 in \textbf{Supplementary Note 3}. The results indicate that our comparison method can distinguish the differences between a real network and the networks generated after certain perturbation.

Figure~\ref{3} shows $D_{NE}$ is an effective way to distinguish networks and shortest path distance based method ($D_{SP}$) can partly tell the difference between different synthetic networks. We further hybridize these two distance distributions to explore the performance of the hybrid method on network comparison.  To recap, we use $P_{N\times N}$ and $H_{N\times L}$ to represent  the shortest path distance distribution and the distance distribution based on network embedding, respectively. As the dimension of $P$ and $H$ is different, we expand short dimension matrix with zero values. That is to say, if $N < L$, we expand $P_i$ to $L\times 1$ dimensions, i.e.,  $P_i=(P_{i1}, P_{i2}, \cdots, P_{iN}, 0, \cdots, 0)$. And if $N > L$, we expand $H_i$ to $N \times 1$ dimensions, i.e., $H_i=(H_{i1}, H_{i2}, \cdots, H_{iL}, 0, \cdots, 0)$. For each node $v_i$, the hybrid distance distribution $M_i$ is defined as the normalization of $\lambda P_i +(1-\lambda)H_i$. Thus, we define the hybrid distance distribution $M$ as

\begin{equation}\label{Dis-M}
        M = \lambda P +(1-\lambda)H,
\end{equation}
where $\lambda$ is a tunable parameter. We use $M$ to replace $H$ in Eq.~(\ref{D-NE}), and obtain the hybrid network comparison method, which is denoted as $D_M$.

\begin{figure}[!ht]
\centering
\includegraphics[width=16cm,height=17cm]{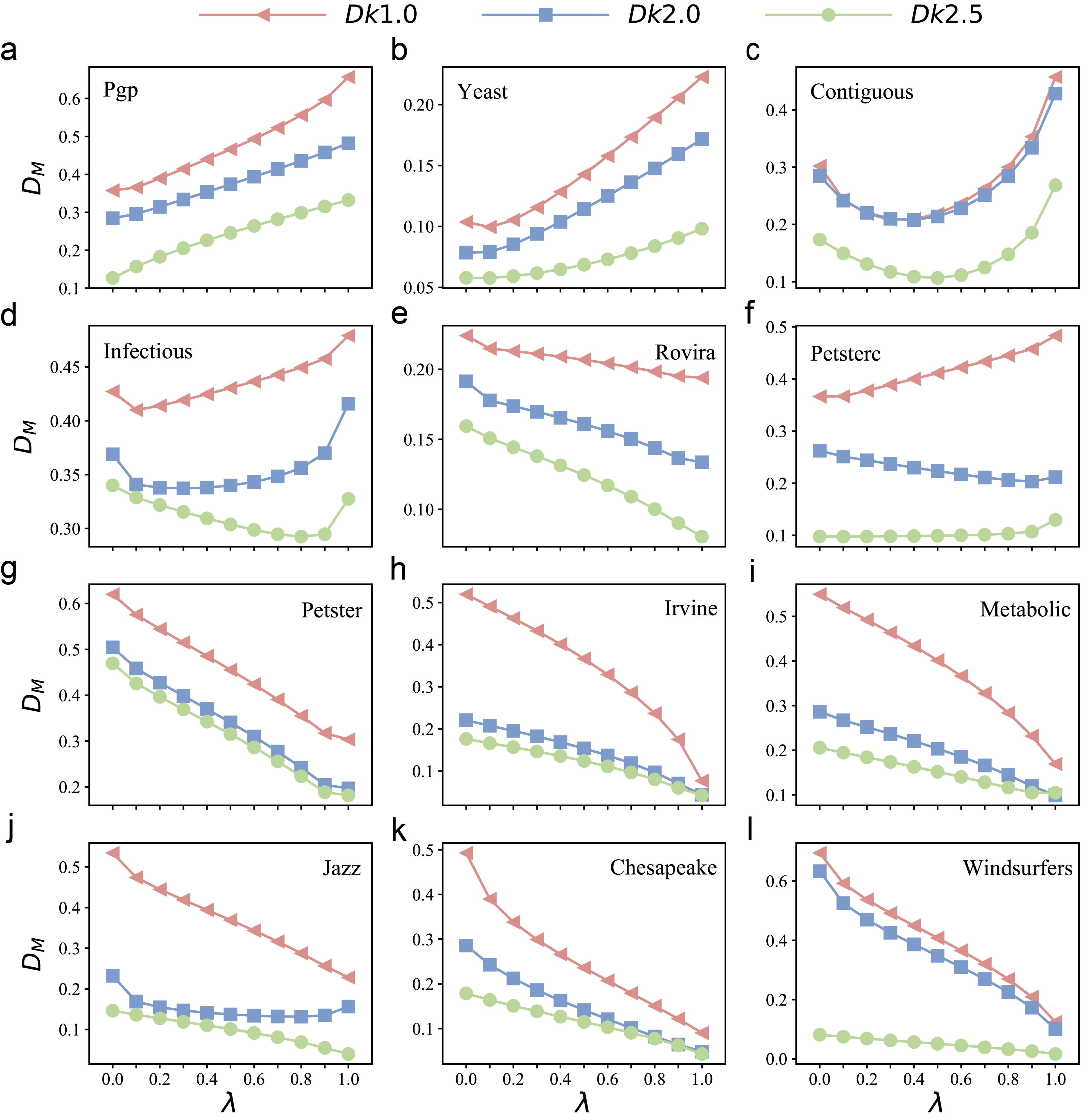}
\caption{The dissimilarity between a network and its null models characterized by the hybrid method.  When parameter $\lambda$=0, the hybrid method degenerates to  $D_{NE}$; when $\lambda$ =1, the real shortest path distance distribution of the network is used to characterize the dissimilarity.}
\label{6}
\end{figure}

We test the performance of $D_M$ on the comparison of a network and its null models (i.e., $Dk1.0$, $Dk2.0$ and $Dk2.5$) in Figure~\ref{6}. We use $D_M(Dk1.0)$, $D_M(Dk2.0)$ and $D_M(Dk2.5)$ to represent the dissimilarity between the original network and its null models, respectively. The pattern of the dissimilarity between a real network and its null models (i.e., $D_M(Dk1.0)>D_M(Dk2.0)>D_M(Dk2.5)$) when $\lambda<1$ is consistent with the order of the  null models. However, $D_M$ can't tell the difference between the network and its null models very well, i.e., $D_M(Dk2.0)$ and $D_M(Dk2.5)$ share the same value in Figure~\ref{6}h, \ref{6}i and \ref{6}k when $\lambda=1$. In fact, $\lambda=1$ indicates that the hybrid distance distribution degrades into only considering the shortest path distance distribution (Eq.~(\ref{Dis-M})), leading to $D_M\approx D_{SP}$ for $\lambda=1$. The network basic features show that the average shortest path length of Irvine (Figure~\ref{6}h), Metabolic (Figure~\ref{6}i), Chesapeake (Figure~\ref{6}k), and  Windsurfers (Figure~\ref{6}l) are significantly smaller than the other networks, which can not be well compared according to $D_M$ for $\lambda=1$. It indicates that $D_{SP}$ can not well tell the difference of the real network with small average shortest path length, which is consistent with the findings in the synthetic networks (Figure~\ref{2}b and \ref{2}e). And for $\lambda=0$, which means the hybrid method degrades into $D_{NE}$, shows better discriminative performance on network comparison across networks with
different average shortest path length, which implies the robustness of $D_{NE}$ upon different network structure.

\begin{figure}[!ht]
\centering
\includegraphics[width=15cm,height=12cm]{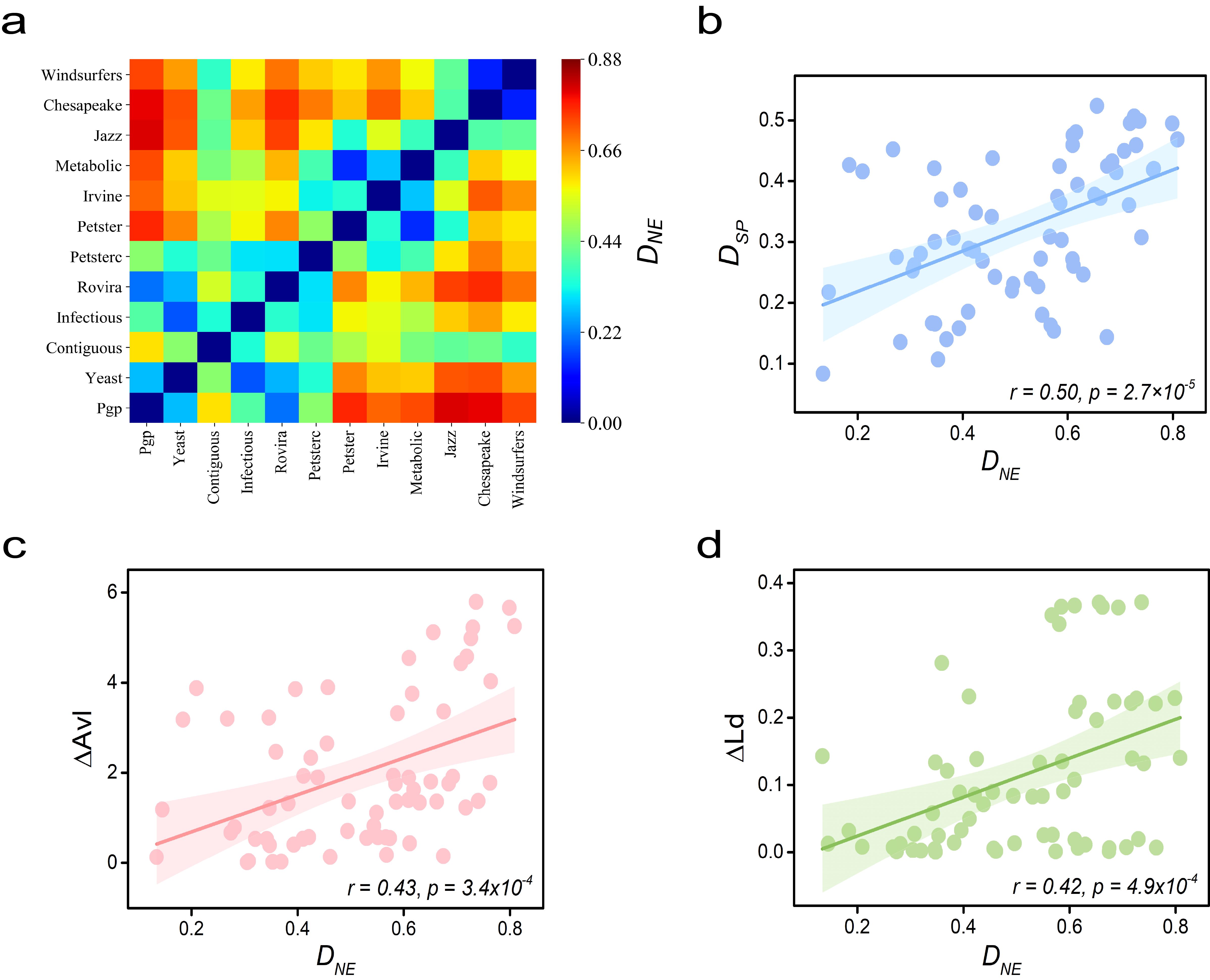}
\caption{(a) Comparison between real networks, in which networks are sorted in descending order based on average shortest path length. (b) Correlation between network comparison methods $D_{SP}$ and $D_{NE}$. (c) Correlation between network dissimilarities $D_{NE}$ and average shortest path length differences $\Delta Avl$ on 12 real networks. (d) Correlation between network dissimilarities $D_{NE}$ and link density differences $\Delta Ld$ on 12 real networks.}
\label{7}
\end{figure}

The dissimilarity between the 12 real networks is given in Figure~\ref{7}. We show $D_{NE}$ between network pairs in Figure~\ref{7}a, we find that networks that have the similar value of average shortest path length tend to be similar.  It implies that $D_{NE}$ considers the path properties of a network when comparing networks. The implication is further amplified by the high Pearson correlation coefficient ($r=0.50, p=2.7\times 10^{-5}$) between $D_{SP}$ and $D_{NE}$ given in Figure~\ref{7}b, where the values of $D_{SP}$ and $D_{NE}$ are computed between the 12 real networks. Given two networks $G_1$ and $G_2$, we define the average shortest path length difference and the link density difference between them as $\Delta Avl$ = $\left|Avl_1 - Avl_2\right|$ and $\Delta Ld$ = $\left|Ld_1 - Ld_2\right|$, respectively. In Figure~\ref{7}c, we show the Pearson correlation between $D_{NE}$ and $\Delta Avl$, which is as high as  $r=0.43$ ($p=3.4\times 10^{-4}$). It further explains the results of Figure~\ref{7}a, i.e., networks with similar average shortest path length tend to be similar.  Meanwhile, the high Pearson correlation coefficient (Figure~\ref{7}d, $r=0.42, p=4.9\times 10^{-4}$) is also found between $D_{NE}$ and $\Delta Ld$. In conclusion, the embedding based network comparison method can capture network properties such as average shortest path length and link density.

\begin{figure}[!ht]
\centering
\includegraphics[width=7.5cm,height=6cm]{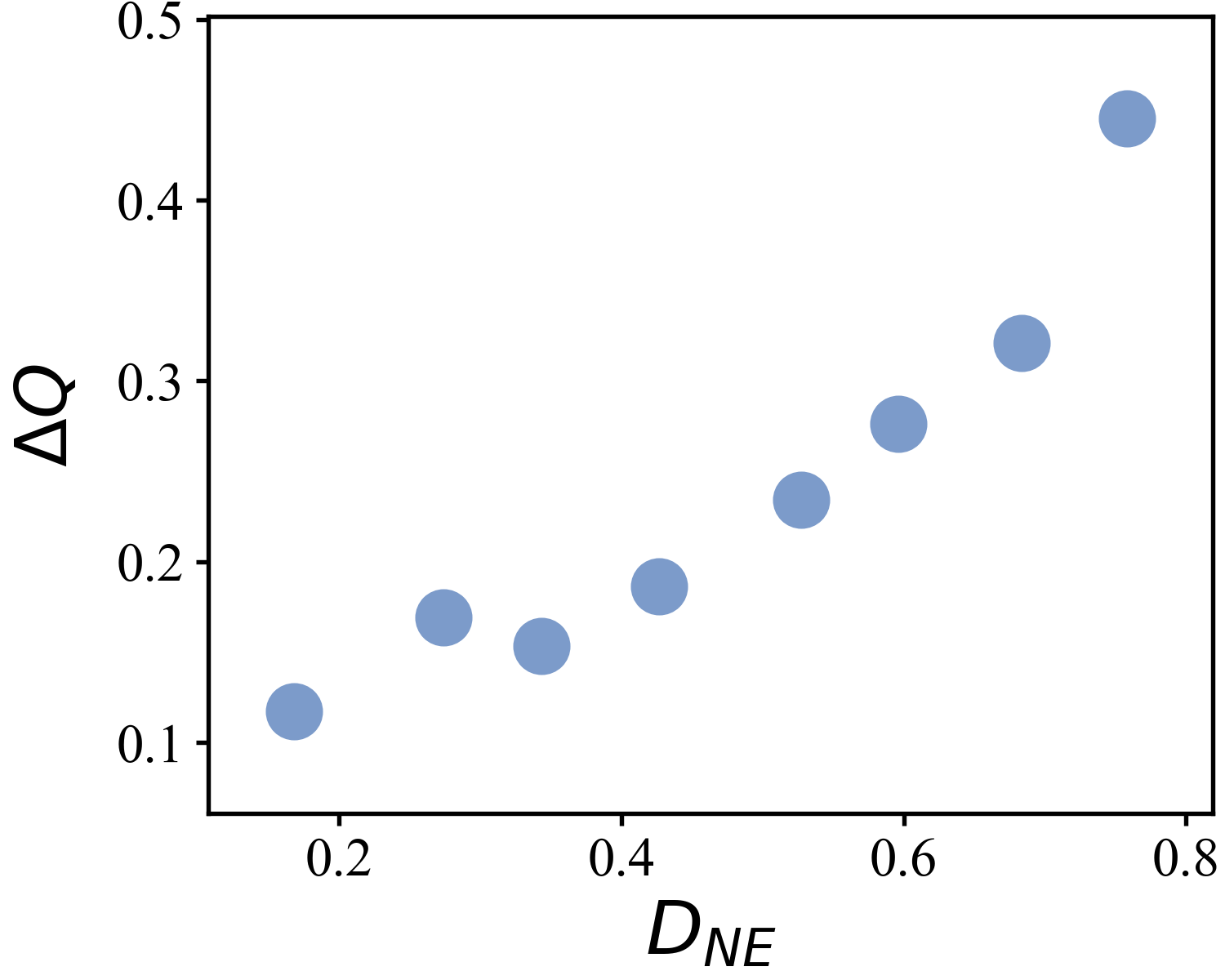}
\caption{Correlation between network dissimilarities $D_{NE}$ and modularity differences $\Delta Q$ on 12 real networks.}
\label{8}
\end{figure}

Modularity reflects the strength of division of a network into communities \cite{newman2006modularity}, i.e., a network with a high modularity indicates that nodes are densely connected within the communities and sparsely connected across different communities. Thus, we explore the relationship between modularity and network structural difference. We define the community segmentation with the maximal network modularity as $Q$, which corresponds to the optimal division of a network \cite{newman2004finding}.  Given two networks $G_1$ and $G_2$, we define the modularity difference between them as $\Delta Q$ = $\left|Q_1 - Q_2\right|$. Figure~\ref{8} shows the correlation between $\Delta Q$ and dissimilarity value $D_{NE}$ on 12 real networks. The result shows that the similar networks tend to have small value of $\Delta Q$ and vice verse. It further emphasizes the good performance of our network embedding based network comparison method.

\section{Conclusions}
In this paper, we propose a network-embedding based network comparison method $D_{NE}$, which is based on node distance distribution and Jensen-Shannon divergence. Specifically, we firstly obtain the embedding vector for each node through \textit{DeepWalk} and calculate the Euclidean distance between each of the node pairs. We measure the distance  distribution heterogeneity of a network via defining the Jensen-Shannon divergence of the node distance  distributions. The dissimilarity between two networks is further defined by the combination of the difference of the average distance  distribution of the networks and the network Euclidean distance  distribution heterogeneity. We compare the proposed method $D_{NE}$ with two state-of-the-art methods, i.e., network dissimilarity based on shortest path distance distribution ($D_{SP}$) and network dissimilarity based on communicability sequence ($D_C$), on various synthetic and real networks. Furthermore, we find that $D_{NE}$  shows better performance in quantifying network difference in almost all the networks. In addition, we find that $D_{NE}$ is also linearly correlated with $D_{SP}$ (Pearson correlation coefficient $r=0.5$), and thus can capture network properties such as average shortest path length and link density. Moreover, it shows that real networks that are similar to each other tend to have small difference in modularity.

We confine ourselves to \textit{DeepWalk} to embed networks, which is a simple and efficient network embedding method. In future work, more advanced embedding methods, such as \textit{Node2Vec} \cite{grover2016node2vec}, graph neural network \cite{zhang2018link}, which may result in more accurate embeddings, could be promising in quantifying network dissimilarity. We deem that our methods can also be generalized to other network types, such as multilayer networks~\cite{kivela2014multilayer}, temporal networks~\cite{holme2015modern}, signed networks~\cite{tang2016survey} and hypergraphs~\cite{bretto2013hypergraph}.

\clearpage

\bibliographystyle{elsarticle-num}
\bibliography{myref}

\end{document}